\documentstyle[emulateapj,psfig]{article}

\newcommand{\bR}{\mbox{\boldmath $R$}}

\begin{document}

\submitted{Accepted for publication in ApJ Letters}
\title{Bar Dissolution in Prolate Halos}
\author{Makoto Ideta}
\affil{
Department of Astronomy, Kyoto University, Kyoto 606-8502, Japan;
  ideta@kusastro.kyoto-u.ac.jp}
\author{and}
\author{Shunsuke Hozumi}
\affil{
Faculty of Education, Shiga University, 2-5-1 Hiratsu, Otsu,
  Shiga 520-0862, Japan; hozumi@sue.shiga-u.ac.jp}

\begin{abstract}

The time evolution of barred structures is examined under the
influence of the external forces exerted by a spherical halo and by
prolate halos.  In particular, galaxy disks are placed in the plane
including the major axis of prolate halos, whose configuration is
often found in cosmological simulations.  $N$-body disks in fixed
external halo fields are simulated, so that bars are formed via
dynamical instability.  In the subsequent evolution, the bars in
prolate halos dissolve gradually with time, while the bar pattern in a
spherical halo remains almost unchanged to the end of the simulation.
The decay times of the bars suggest that they can be destroyed in a
time smaller than a Hubble time.  Our results indicate that this
dissolution process could occur in real barred galaxies, if they are
surrounded by massive dark prolate halos, and the configuration
adopted here is not unusual from the viewpoint of galaxy formation.
For a prolate halo model, an additional simulation that is restricted
to two-dimensional in-plane motions has also ended up with similar bar
dissolution.  This means that the vertical motions of disk stars do
not play an essential role in the bar dissolution demonstrated here.

\end{abstract}

\section{Introduction}

Observations of disk galaxies have shown that their rotation curves
are often flat out to large distances (e.g., \cite{rftb82}, 1985;
\cite{smk87}).  This finding suggests that disk galaxies, regardless
of whether they are barred or non-barred, are surrounded by massive
dark halo.  However, it is difficult to distinguish between barred and
non-barred galaxies from edge-on views which can easily provide
rotation curve data.  Recent development of observations enables us to
measure rotation curves of barred galaxies, even though they are not
viewed fully edge-on.  As a result, it has been confirmed that many,
if not all, barred galaxies show flat rotation curve, which does
suggest the existence of massive dark halo in barred galaxies as well
(\cite{jk83}; \cite{pffma94}).

The shape of dark matter halo, on the other hand, remains uncertain
observationally, although some indication is obtained (\cite{pds99}).
Cosmological simulations demonstrate that dark matter halos are
inclined to be more frequently prolate than oblate (\cite{dc91};
\cite{wqsz92}).  In addition, prolate halos can be supported from the
standpoint of the longevity of galactic warps; Ideta et al.\ (2000)
have shown that a warped structure persists for a long time in a
prolate halo while it disappears quickly in an oblate halo.

Recently, El-Zant \& Ha\ss ler (1998) have revealed that bars are
likely to be destroyed owing to chaotic diffusion of bar-supporting
orbits, if they are embedded in triaxial halos including prolate
configurations.  Their results indicate that such bar destruction is
conceivable from the situation where a barred galaxy is placed in a
prolate halo with the major axis of the halo being in the disk plane.
According to Warren et al.'s (1992) cosmological simulations, this
configuration of a disk in a prolate-like halo often occurs.  Thus,
real barred galaxies might be surrounded by prolate halos, and could
suffer forces caused by the elongated potential of those halos.
Unfortunately, however, El-Zant \& Ha\ss ler (1998) treated all the
components of a bar, disk and halo as fixed potentials, and analyzed
the orbits of test stars.  Consequently, it is unclear whether bars
are in reality destroyed in such prolate halos.

In this {\it Letter}, we demonstrate from $N$-body simulation that a
barred structure is, in fact, completely destroyed in a time smaller
than a Hubble time if it is embedded in a prolate halo whose major
axis is placed in the disk plane.  In \S 2, we describe the initial
setup of our models and the numerical method.  Results are presented
in \S 3.  A discussion and conclusions are given in \S 4.

\section{Models and Method}

We set up a configuration such that the mid-plane of a galaxy disk is
put in the plane including the major axis of a prolate halo.  For
comparison, a spherical halo is also included in our models.  As a
first step, the halo is handled as a fixed external field.  We will
discuss the possible effects of a live halo on the evolution of a bar
pattern in the last section.  The disk is evolved forward in time to
form a bar via dynamical instability.  Thereafter, we examine how the
barred structure is influenced in the prolate halo.  Since we pay
attention to mainly the effects of an elongated halo on a bar, we do
not include a bulge component which acts to reduce the amplitudes of
the bars arising from the bar instability (\cite{hfn87}).

The prolate halo models are constructed from an axisymmetric
modification of Hernquist's models (\cite{lh90}).  The density profile
of the halos is represented by
\begin{equation}
  \rho_{\rm h}\left(m\right)=\frac{M_{\rm h}}{2\pi ac^2}\frac{1}
  {m\left(1+m\right)^3},
\end{equation}
where $M_{\rm h}$ is the halo mass, $a$ and $c$ are the scale lengths
along the major and minor axes, respectively, and
\begin{equation}
  m^2=\frac{x^2}{a^2}+\frac{y^2+z^2}{c^2}.
\end{equation}
If the ratio of $c$ to $a$ is smaller than unity, the halo is prolate.
We choose $c/a=0.6$ and 0.75 as prolate halos, and $c/a=1$ as a
spherical halo.  The models with $c/a=0.6, 0.75$, and 1 are named
Models P060, P075, and S100, respectively.

We adopt an exponential surface-density profile (\cite{kcf70}) with an
isothermal density distribution in the vertical direction
(\cite{ls42}) for the disk models that are given by
\begin{equation}
  \rho_{\rm d}\left(R,z\right)=\frac{M_{\rm d}}{4\pi R^2_{\rm d}z_{\rm d}}
  \exp\left(-\frac{R}{R_{\rm d}}\right){\rm sech}^2\left(\frac{z}{z_{\rm d}}
  \right),
\end{equation}
where $M_{\rm d}$ is the disk mass, $R_{\rm d}$ is the scale length,
and $z_{\rm d}$ is the scale height.  The disks are truncated radially
at $15\;R_{\rm d}$, and vertically at $2\;z_{\rm d}$.  The velocity
distribution of disk particles is realized by employing Hernquist's
(1993) approach which is based on the moments of the collisionless
Boltzmann equation.  In setting up the particle velocities, the disk
plane is placed to include the major axis of the prolate halo.  The
typical Toomre $Q$ parameter (\cite{at64}) is chosen to be of order
unity so as to induce the bar instability easily, and the value,
$Q=1$, is set at the radius corresponding to the sun in the physical
units of the Galaxy described below.

\begin{center}
{\footnotesize
TABLE~1 \\
{\sc Parameters for $N$-body Simulations}
{\vskip .4em}}
{\footnotesize
\begin{tabular}{ccccccc}
\tableline
\tableline
\noalign{\vskip .4em}
Model & $M_{\rm d}$ & $R_{\rm d}$ & $z_{\rm d}$ & $M_{\rm h}$ & $a$ &
$c$ \\
\tableline
\noalign{\vskip .4em}
P060..... & 1.0 & 1.0 & 0.2 & 9.0 & 10.0 &  6.0 \\
P075..... & 1.0 & 1.0 & 0.2 & 9.0 & 10.0 &  7.5 \\
S100..... & 1.0 & 1.0 & 0.2 & 9.0 & 10.0 & 10.0 \\
\tableline
\end{tabular}}
\end{center}

\medskip
\centerline{\psfig{figure=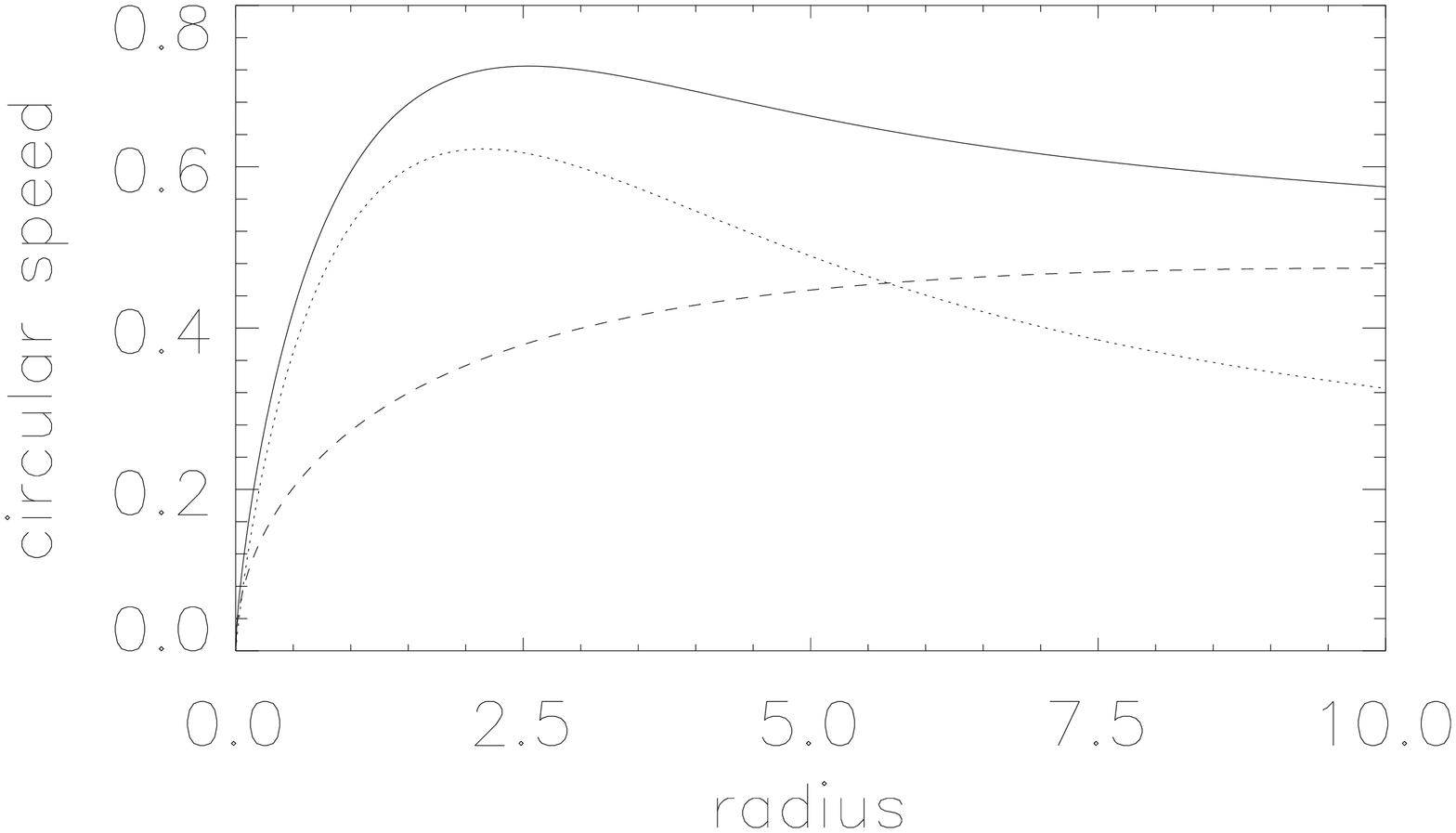,width=2.00in,angle=90}}
\figcaption{
Circular speed showing the contribution from the halo (dashed line)
and disk (dotted line) to the total (solid line) for Model S100.
\label{fig:cirvel}}
\medskip

We employ a system of units such that the gravitational constant
$G=1$, the disk mass $M_{\rm d}=1$, and the scale length $R_{\rm
d}=1$.  If these units are scaled to physical values appropriate for
the Milky Way, i.e., $R_{\rm d}=3.5\;{\rm kpc}$ and $M_{\rm d}=5.6
\times 10^{10}\; M_\odot$, unit time and velocity are
$1.31\times10^7\;{\rm yr}$ and $262\;{\rm km\;s^{-1}}$, respectively.
The scale height $z_{\rm d}$ is set to be $0.2$, or in our units,
$700\;{\rm pc}$, which corresponds to that of old stars in the Milky
Way.  The disk is represented by 100,000 particles of equal mass.
Since flat rotation curves of spiral galaxies suggest that the mean
halo-to-disk mass ratio at the Holmberg radius is 1.0 (e.g.,
\cite{cf85}), we determine the halo mass so that the disk and halo
masses within $5\;R_{\rm d}$ are equal to each other in the spherical
halo model.  The parameters of each model are presented in Table 1,
and the rotation curve of Model S100 is shown in Figure
\ref{fig:cirvel}.

We use a hierarchical tree algorithm (\cite{bh86}) with an opening
angle criterion, $\theta=0.75$, being adopted.  We expand forces and
potentials up to quadrupole terms in the tree code.  Forces are
softened with a cubic spline (\cite{hk89}), and the spline softening
length is $0.04\;R_{\rm d}$, or in other words, $0.2\;z_{\rm d}$.  The
equations of motion are integrated with a fixed timestep, $\Delta
t=0.04$, using a leapfrog method.  For these choices of parameters,
the total energy was conserved to better than $0.13$\% in all
simulations.

\section{Results}

\begin{figure*}[ht]
\centerline{\psfig{figure=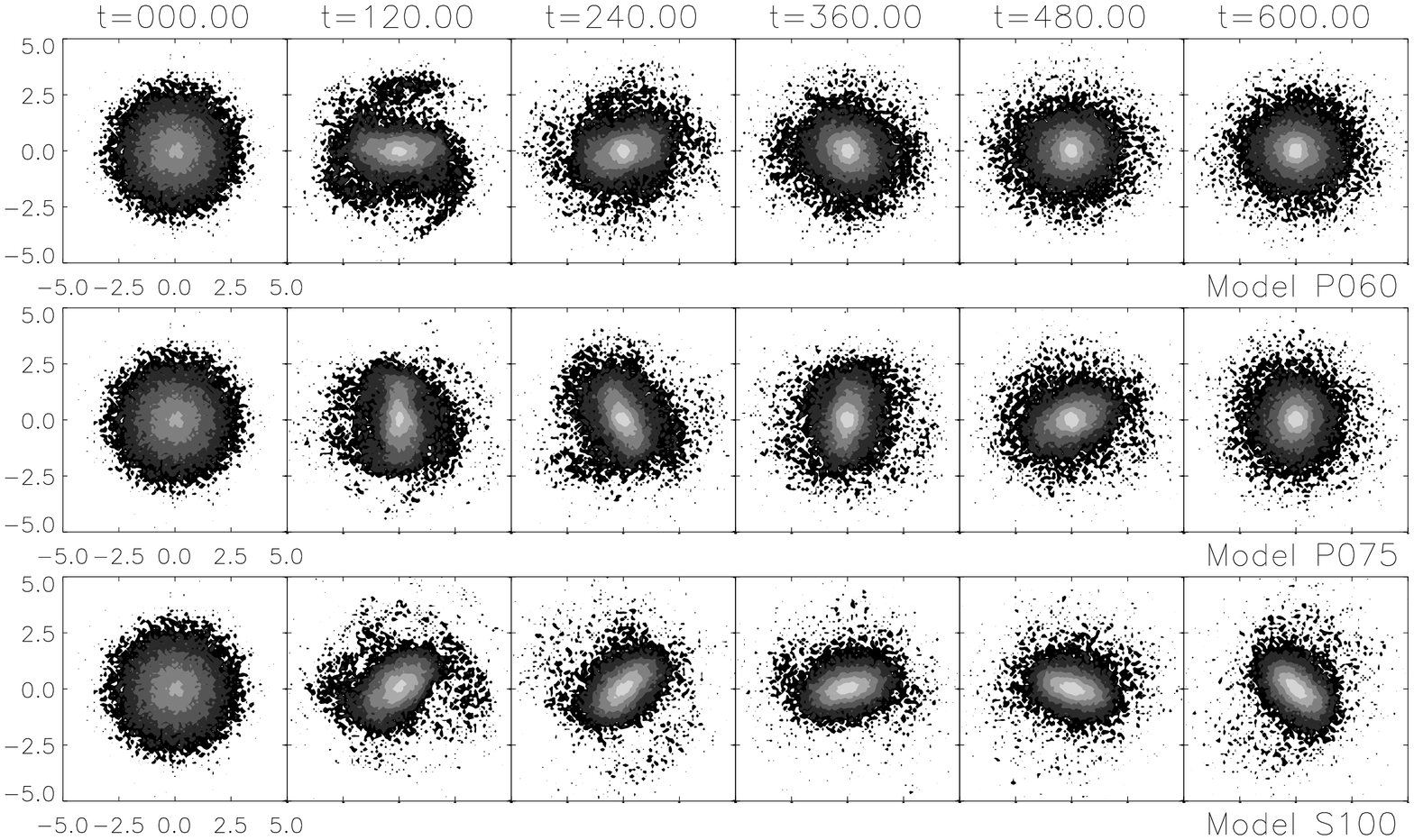,width=5.00in,angle=90}}
\figcaption{Time evolution of the face-on density contours for Models
P060 (top row), P075 (middle row), and S100 (bottom row).  In all the
models, bars are almost fully developed by t=120.  The bar in the
spherical halo model (Model S100) continues to exist to the end of the
simulation, while the bars in the prolate halo models (Models P060 and
P075) are weakened gradually with time and are completely destroyed at
the end.
\label{fig:faceon}}
\end{figure*}

Figure \ref{fig:faceon} shows the contour plots of face-on density
profiles at intervals of 120 time units for Models P060, P075, and
S100.  In all the models, bars were formed through the bar
instability, and they were almost fully developed by $t\simeq 120$.
We can see from Figure \ref{fig:faceon} that in the subsequent
evolution, the bar pattern was kept nearly unchanged to the end of the
simulation for the spherical halo model (Model S100), while the bars
continued to dissolve gradually with time for the prolate halo models
(Models P060 and P075).  To evaluate the change in bar shape, we
derived the axis ratio of the bars from the moment of inertia tensor
for disk particles included in $2.0\;R_{\rm d}$ at which the bars end
roughly.  Then, we have found that the axis ratio changed from $\simeq
0.71$ at $t=150$ to $\ga 0.95$ at $t=600$ for Model P060, and that it
did from $\simeq 0.68$ at $t=150$ to $\ga 0.93$ at $t=600$ for Model
P075.

To quantify the deformation of the bars formed in each halo model, we
calculated their amplitudes as follows.  The bars are very thin as
compared to the extent of the disks, so that we ignore the thickness
of the bars by projecting the particle distributions to the mid-plane
of the disks.  Then, we expand the density and potential of the
projected distribution in a set of basis functions as
\begin{eqnarray}
\mu\left(\bR\right)=\sum_{n,m}A_{nm}\mu_{nm}\left(\bR\right),\\
\Phi\left(\bR\right)=\sum_{n,m}A_{nm}\Phi_{nm}\left(\bR\right),
\end{eqnarray}
where each density-potential pair, $\mu_{nm}$ and $\Phi_{nm}$, constitutes
Aoki \& Iye's (1978) basis set given by
\begin{eqnarray}
\label{eq:aimu}
\mu_{nm}\left(\bR\right)=\frac{2n+1}{2\pi}\left(\frac{1-\xi}{2}
  \right)^{3/2}P_{nm}\left(\xi\right)\exp\left(im\theta\right), \\
\label{eq:aiphi}
\Phi_{nm}\left(\bR\right)=-\left(\frac{1-\xi}{2}\right)^{1/2}
  P_{nm}\left(\xi\right)\exp\left(im\theta\right).
\end{eqnarray}
Here, $P_{nm}$ are the Legendre functions ($n\ge m$), and the radial
transformation
\begin{equation}
\xi=\frac{R^2-1}{R^2+1}
\end{equation}
is used.  Positive values of $m$ correspond to the number of arms in
spiral patterns.  In the expansions shown above, the amplitude of the
$(n,m)$-mode is calculated from the absolute value of the expansion
coefficient, $|A_{nm}|$.  In particular, we pay attention to the
fastest growing bar mode with $(n,m)=(2,2)$.

\medskip
\centerline{\psfig{figure=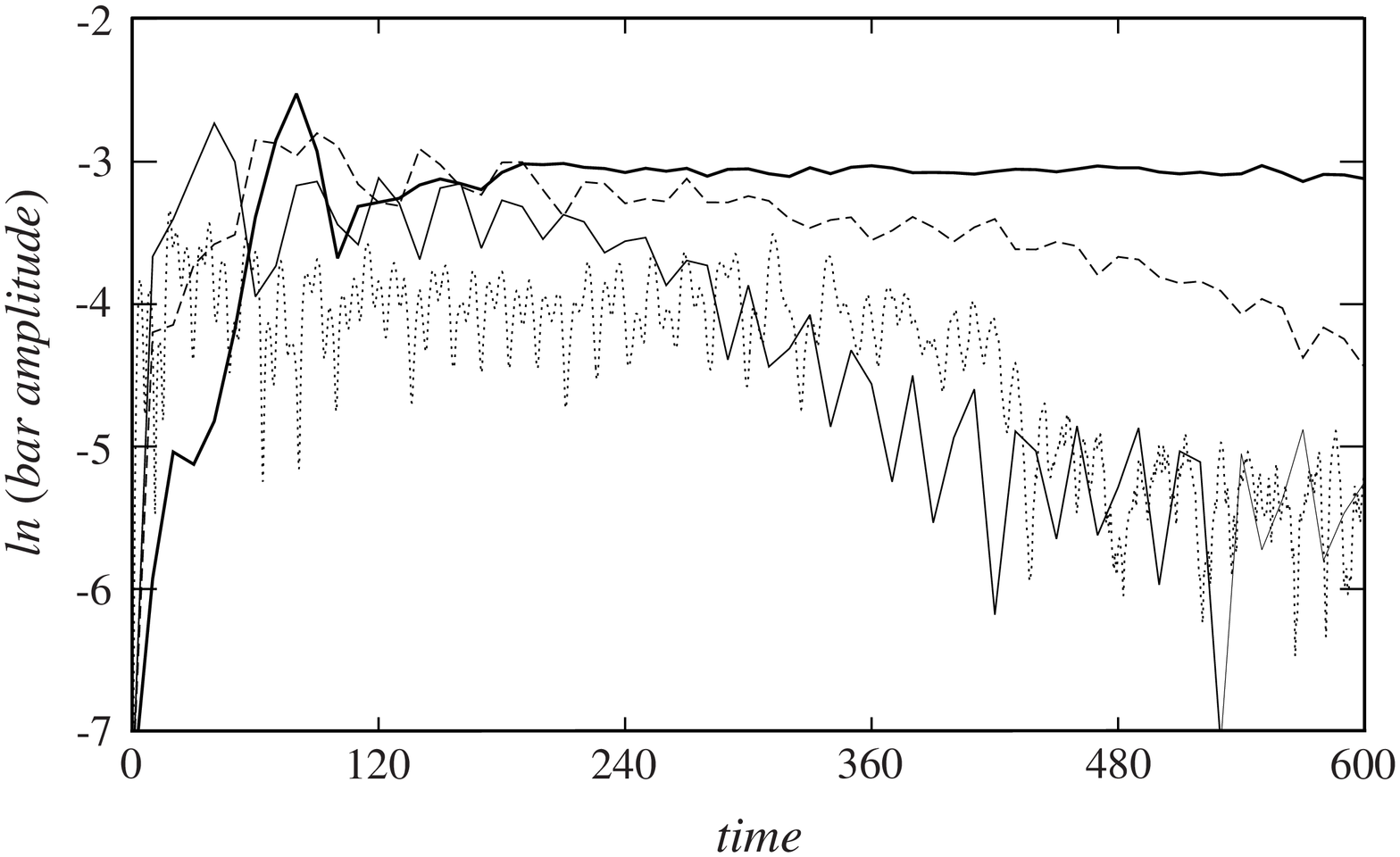,width=3.25in,angle=90}}
\figcaption{
Time evolution of the bar amplitude of the fastest growing mode for
Models P060 (thin solid line), P075 (dashed line), and S100 (thick
solid line).  Also shown is the time evolution of the bar amplitude
for a two-dimensional analogue of Model P060 (dotted line).
\label{fig:baramp}}
\medskip

In Figure \ref{fig:baramp}, we present the time evolution of the bar
amplitude, $|A_{22}|$, for each halo model.  This figure describes the
behavior of the barred structures seen in Figure \ref{fig:faceon}
quantitatively.  Again, we find that the bars have grown up to their
full amplitudes by $t\simeq 120$.  In particular, it should be
emphasized that the bar amplitude decays nearly exponentially with
time $\sim \exp(-t/\tau)$ for the prolate halo models while it remains
practically constant to the end of the simulation for the spherical
halo model.  To estimate the decay times $\tau$, least-squares fits
were applied to the data from $t=200$ to $t=400$ for Model P060, and
to those from $t=200$ to $600$ for Model P075.  We obtained
$\tau\simeq 111$ and $\tau\simeq 375$ for Models P060 and P075,
respectively.  If these values are represented by the physical units
appropriate for the Milky Way, the decay times correspond to $\sim
1.5\times 10^9$ yr and $\sim 4.9\times 10^9$ yr, respectively.  These
time intervals are relatively smaller than a Hubble time that is
almost equivalent to the supposed ages of disk galaxies, even though
it takes about twice as long as the decay times for the bar amplitude
to decrease by an order of magnitude.  Thus, this kind of bar
dissolution addressed here could occur in real barred galaxies,
provided that our models reflect physical realism to a reasonable
degree.

\section{Discussion and Conclusions}

We have found that a prolate halo can destroy a bar in a time smaller
than a Hubble time, if the major axis of the halo lies in the disk
plane.  It has been known that a bar can be destructed by central
massive objects (\cite{nsh96}; \cite{hh99}), and also by close
encounter or minor merger of nearby dwarf galaxies (\cite{ea96}).  Our
finding adds another situation where bar dissolution can occur even
without recourse to massive central objects or nearby companion
galaxies.  Thus, in considering the evolution of barred galaxies, we
should take into account the shape of surrounding halos and their
configurations to the disks that support the bars.

The time scale of bar dissolution will depend on the parameters of
halos such as the mass, core radius, and axis ratio.  Then, to
investigate whether real barred galaxies can experience bar
dissolution, we need to check out how realistic our models are.
First, for a halo model, we have used Hernquist's models that have a
density profile proportional to $r^{-1}$ near the center.  Such a
central cuspy density profile might be a natural end-product of
hierarchical clustering based on a standard cold dark matter scenario
(e.g., \cite{nfw96}, 1997; \cite{fm97}).  From this point of view, our
adopted halo density profiles would not be peculiar.  Next, the halo
mass and core radius were chosen to follow the observed rotation
curves in spiral galaxies such that the circular velocities are almost
constant out to large radii.  Thus, the kinematic structures of the
systems adopted here would not differ substantially from those of real
galaxies.  Last, Model P075 $(c/a=0.75)$ also leads to bar destruction
within a Hubble time.  Since cosmological simulations demonstrate that
the angular momentum vector of galactic disks is often aligned to the
minor axis of dark halos, the axis ratio of halos in the disk plane
corresponds to the major-to-intermediate one.  According to Figure 7a
of Warren et al.\ (1992), the cumulative fraction of the dark halos
that have a major-to-intermediate axis ratio smaller than 0.75 amounts
to about 70\%.  In addition, the observationally determined
geometrical form of dark halos is shown to be $(b/a)_\rho\ga 0.8$ as
the equatorial axis ratio in density, and $(c/a)_\rho=0.5\pm 0.2$ as
the vertical-to-equatorial axis ratio in density (\cite{pds99}).
These values indicate that halos tend to be prolate with the major
axis being in the disk plane, and that the axis ratio of $c/a=0.75$
employed in our simulation is not far from the current lowest limit of
0.8.  Taking into consideration those aspects mentioned above, bar
dissolution might be in progress in the real universe.

As a possible mechanism of bar destruction, Raha et al.\ (1991)
pointed out the bar buckling instability that is known as the firehose
instability.  This instability originates in the anisotropic velocity
dispersion between the motions parallel to and those vertical to the
disk.  If this is the case for the bar destruction presented here, a
bar should survive in an infinitesimally thin disk, irrespective of
the surrounding halo shape.  Then, we run an additional simulation in
which the motions of stars are restricted to the plane of the disk
with a self-consistent field (SCF) method (\cite{mcb72}; \cite{ho92})
using Aoki \& Iye's basis set (eqs.\ [\ref{eq:aimu}] and
[\ref{eq:aiphi}]).  The model realized with 100,000 particles of equal
mass is a two-dimensional analogue of Model P060.  In the SCF code,
the maximum number of radial expansion coefficients is taken to be 16,
and that of azimuthal ones is chosen to be 2 with only even values
being used ($m=0$, and 2).  The resulting behavior of the bar
amplitude defined by $|A_{22}|$ is indicated by the dotted line in
Figure \ref{fig:baramp}.  Clearly, the bar amplitude decreases nearly
exponentially with time, and its time evolution is quite similar to
that in the corresponding three-dimensional simulation (Model P060).
This fact implies that the buckling instability should not be the
essential cause of the bar destruction found in our three-dimensional
simulations.

Another likely mechanism of bar destruction is the effect of chaotic
orbits.  Recently, El-Zant \& Ha\ss ler (1998) have found that the
orbits that support a bar are highly chaotic when the barred galaxy is
embedded in a triaxial halo whose minor axis is vertical to the disk
plane.  Then, they suggest that the bar may be destroyed owing to
chaotic diffusion, although they do not mention the destruction time
scale.  Norman et al.\ (1996) have revealed that a bar is destroyed
within a few dynamical times after the contraction of a rigid mass
component, whose mass exceeds some critical value, is completed to
form a central massive object.  As a result, they are led to a picture
that the abrupt destruction of a bar results from the chaotic behavior
of bar-supporting orbits.  However, our results show the gradual
erosion of bars, and so, the behavior of the bar amplitude is closer
to that found by Hozumi \& Hernquist (1999) who examined secular
change in bar pattern caused by central massive black holes added to
flat disks.  Of course, chaotic orbits might lead to the gradual
destruction of bars.  Therefore, we will analyze our models in phase
space to study stochasticity in stellar orbits in a separate paper.

Our simulations are highly idealized in that halos were treated as
external fixed potentials.  If halos are made mobile, the pattern
speed of a bar will quickly slow down owing to dynamical friction of
the halos (\cite{mdw85}; \cite{ds98}).  It is conceivable that the
time scale of bar destruction could be affected by the pattern speed
of a bar.  Thus, a wider variety of simulations with a greater degree
of realism will be required to figure out the detailed processes of
bar dissolution.

\acknowledgements

We are grateful to Dr.\ T.\ Tsuchiya for useful discussions.  We also
thank Dr.\ K.\ Ohta for helpful comments on the observational side of
barred galaxies.  Numerical simulations were carried out on VPP/16R
and VX/4R at the Astronomical Data Analysis Center of the National
Astronomical Observatory, Japan (ADAC/NAOJ).

\end{document}